\documentclass[12pt,letterpaper,nohyper]{JHEP3}

\usepackage{amsmath,amssymb,amsfonts,epsfig,slashed,xspace
}

\title{N=4 BPS black holes and octonionic twistors}

\author{Yann Michel$^1$, Boris Pioline$^{1,2}$ and Cl\'ement Rousset$^1$ \\
$^1$ Laboratoire de Physique Th\'eorique et Hautes
Energies\footnote{Unit\'e mixte de recherche du CNRS UMR 7589},\\
Universit\'e Pierre et Marie Curie - Paris 6,
4 place Jussieu, F-75252 Paris cedex 05 \\

$^2$ Laboratoire de Physique Th\'eorique de l'Ecole Normale
Sup\'erieure\footnote{Unit\'e mixte
de recherche du CNRS UMR 8549},\\
24 rue Lhomond, F-75231 Paris cedex 05\\
\\
{\tt E-mail: yann,pioline,rousset@lpthe.jussieu.fr}}

\preprint{LPTENS 08/36}

\abstract{Stationary, spherically symmetric solutions of $\cN=2$ supergravity
in 3+1 dimensions have been shown to correspond to holomorphic curves
on the twistor space of the quaternionic-K\"ahler space which arises in the
dimensional reduction along the time direction. In this note, we 
generalize this result to the case of $1/4$-BPS black holes in $\cN=4$
supergravity, and show that they too can be lifted to holomorphic curves
on a "twistor space" $Z$, obtained by fibering the Grassmannian $F=SO(8)/U(4)$ 
over the moduli space in three-dimensions $SO(8,n_v+2)/SO(8)\times SO(n_v+2)$. 
This provides a kind of octonionic generalization 
of the standard constructions in quaternionic geometry, and may be useful 
for generalizing the known BPS black hole solutions, and finding new non-BPS 
extremal solutions.}

\def\bea{\begin{eqnarray}}
\def\eea{\end{eqnarray}}
\def\be{\begin{equation}}
\def\ee{\end{equation}}
\def\ba{\begin{align}}
\def\ea{\end{align}}
\def\bse{\begin{subequations}}
\def\ese{\end{subequations}}

\newcommand{\pa}{\partial}

\newcommand{\nn}{\nonumber}
\newcommand{\eps}{\epsilon}
\newcommand{\IR}{\mathbb{R}}
\newcommand{\IC}{\mathbb{C}}
\newcommand{\CI}{\mathbb{I}}

\newcommand{\cN}{\mathcal{N}}
\newcommand{\cM}{\mathcal{M}}
\newcommand{\tzeta}{{\tilde\zeta}}
\newcommand{\txi}{{\tilde\xi}}
\newcommand{\vareps}{{\varepsilon}}
\newcommand{\bX}{{\bar X}}
\newcommand{\mX}{\mathcal{X}}
\newcommand{\mXt}{\widetilde{\mathcal{X}}}
\newcommand{\mZ}{\mathcal{Z}}
\newcommand{\qk}{quaternionic-K\"ahler }

\begin{document}

\section{Introduction}

While supersymmetry often leads to solvability, its full power reveals itself only  
when translated into holomorphy. For supergravity theories with $\cN=2$ supersymmetries
in 4 dimensions, this may be achieved using projective superspace \cite{Karlhede:1984vr} 
or harmonic superspace techniques  \cite{Galperin:1984av}. From a mathematical viewpoint,
these techniques are closely related to twistors, whose purpose is to enforce holomorphy in 
all complex structures at once. While these methods have often been used to restrict 
the possible terms in the low energy effective action, they can also be useful in
constructing actual supersymmetric solutions of the field 
equations  \cite{Neitzke:2007ke,Lopes Cardoso:1998wt,Gunaydin:2007bg,Gaiotto:2007ag}.

In particular, in \cite{Neitzke:2007ke,Gunaydin:2007bg,Pioline:2006ni}, it was shown that spherically 
symmetric BPS black hole solutions in $\cN=2$ supergravity correspond to holomorphic 
curves in $Z$, the twistor space of the \qk moduli space $\cM_3$ which appears after dimensional 
reduction along the time direction. This translation of supersymmetry to holomorphy was
then used to recover the known spherically symmetric BPS solutions, and to obtain the
exact quantum wave function for the radial evolution of the scalar fields, at two derivative 
order. It is likely that multi-centered BPS solutions could also be understood or
generalized using the same geometrical framework.  
 
The purpose of this note is to extend the techniques of  \cite{Neitzke:2007ke,Gunaydin:2007bg}
to the case of $\cN=4$ supergravity with $n_v$ vector multiplets in 3+1 dimensions. Since the moduli space for such theories in three-dimensions is a symmetric space $\cM_3=K_3\backslash G_3=
SO(8)\times SO(n_v+2)]\backslash SO(8,n_v+2)$
\cite{Marcus:1983hb,Sen:1994wr}, spherically symmetric solutions can be 
readily obtained by exponentiating a one-parameter subgroup and so hold little mystery.
Nevertheless, with a view to a possible extension to the multi-centered case, or to 
the inclusion of  higher derivative corrections such as the one uncovered 
in \cite{Antoniadis:2006mr}, it is interesting to see how the translation of supersymmetry 
to holomorphy takes place. 

While an approach based on harmonic superspace
ideas is also possible  \cite{Antoniadis:2007cw}, we prefer to follow the road of projective 
superspace, and the guidance of 1/4-BPS black holes. By including the pair of Killing spinors
preserved by the solution into the phase space of the dynamical system governing
the radial evolution equations, we show that  BPS solutions can again be lifted to 
holomorphic curves in the "twistor space" $Z=M_3\backslash G_3=[U(4)\times 
SO(n_v+2)]\backslash SO(8,n_v+2)$, whose fiber $F$ over any point in $\cM_3$
is the Grassmanniann $U(4)\backslash SO(8)=[SO(2)\times SO(6)]\backslash SO(8)$.
The twistor space $Z$ appears in Bryant's classification of twistor spaces of symmetric spaces \cite{Bryant}, and its relevance for black holes was first suggested in \cite{Gunaydin:2007bg}.
In contrast to the standard twistor space for \qk manifolds, $Z$ does not have a (twisted)
holomorphic contact form, but instead an antisymmetric $4\times 4$ matrix of them, 
transforming into each other under the local $SU(4)$ action. This complication
prevents us from constructing a complex coordinate system adapted to the Heisenberg
group of symmetries which is crucial for applications to black holes, although there is
little doubt that such a system exists. Similarly, we fail to produce the most general
black hole wave function, but we do exhibit some holomorphic wave functions.   

The outline of this note is as follows. In Section 2, we review the equivalence between
stationary, spherically symmetric solutions in 4D and geodesic motion in 3D,
derive the supersymmetry conditions, and obtain BPS and non BPS solutions
by exponentiating one-parameter subgroups in $G_3$. In Section 3, we construct
the twistor space $Z$, first in a "bottom-up" approach suggested by the black 
hole problem, and second in a more algebraic "top-down" approach analogous
to the construction in \cite{Gunaydin:2007qq}. The equivalence
between BPS solutions and holomorphic curves is explained in Section 3.4. 
In the appendices, we state our conventions for $SO(8)$ Dirac matrices,
and review some general facts about nilpotent co-adjoint orbits in orthogonal
groups.

\section{Black holes and Geodesics}

\subsection{$\cN=4$ supergravity in four dimensions}

Consider $\cN=4$ supergravity in 3+1 dimensions with $n_v$ vector
multiplets \cite{Cremmer:1977tt,Maharana:1992my}. The spectrum consists 
of the graviton, 4 gravitini, 
$n_v+6$ Abelian vector fields, $n_v+1$ Majorana spinors and  $6n_v+2$ real
scalar fields parametrizing the moduli space 
\be 
\label{m4}
{\cal M}_4=\frac{Sl(2,\IR)}{U(1)}\times\frac{SO(6,n_v,\IR)}
{SO(6)\times SO(n_v)}\ ,
\ee
The first factor in \eqref{m4} corresponds to the axion-dilaton field
$\tau=\tau_1+i\tau_2$ from the gravity supermultiplet, while the
second factor corresponds to the scalars in the $n_v$ vector multiplets. 
The $U(1)$ and $SO(6)$ subgroups in the denominator of \eqref{m4}
correspond to the R-symmetry group $U(4)$.

An $\cN=4$ supergravity with $n_v=22$ vector multiplets is known to arise  
by toroidal compactification of the heterotic string on $T^6$. Theories with
fewer vector multiplets can be constructed by freely acting orbifolds of
this model \cite{Chaudhuri:1995fk}. In these cases, as long as $n_v\geq 6$, it is convenient 
to parametrize the second 
factor of  \eqref{m4} by the coset element
\be
\label{e6nviwa}
e_{6,n_v}=\begin{pmatrix}e_6 & 0 & 0\\0 & \CI_{n_v-6} & 0\\ 0 & 0 &
e_6^{-T}\end{pmatrix}.\begin{pmatrix}\CI_6 & W & B-\frac{1}{2}W^T 
\eta_{6,n_v} W\\0 &
\CI_{n_v-6} & -W^T \eta_{6,n_v} \\ 0 & 0 & \CI_6\end{pmatrix}\,
\in SO(6,n_v,\mathbb{R})
\ee 
which preserves the signature $(+_6,-_{n_v})$ metric
\be
\eta_{6,n_v}=\begin{pmatrix}&&\CI_6\\&-\CI_{n_v-6}&\\\CI_6&&\end{pmatrix}
\ee
Here, $e_6\in\frac{GL(6,\mathbb{R})}{SO(6)}$ is the viel-bein for the metric
on $T^6$, which can be chosen in upper triangular form, $B$ is an antisymmetric
$6\times 6$ matrix corresponding to the Kalb-Ramond two-form pulled back to $T^6$,
and $W$ is a $6\times (n_v-6)$ matrix corresponding to the Wilson lines of the 
$n_v$ Abelian gauge fields in the Cartan subgroup of the 10D gauge 
group (or its projection in the case of CHL compactifications). 
When $n_v< 6$, one may instead use the 
decomposition of $\mathfrak{so}(6,n_v,\mathbb{R})$ as the sum of a compact
(i.e. antisymmetric) and a non-compact (symmetric) element, and
parametrize the second factor in \eqref{m4} by
a real $6\times n_v$-matrix $A$,
\be
\label{e6nvas}
e_{6,n_v}=\exp\begin{pmatrix}0_6&A\\A^T&0_{n_v}
\end{pmatrix}\ ,\quad 
\eta_{6,n_v}=\begin{pmatrix}\CI_6&\\&-\CI_{n_v}\end{pmatrix}\ .
\ee
For type II compactifications on $K_3\times T^2$, or freely-acting orbifolds
thereof, other parametrizations 
adapted to the $SO(4,20)$ mirror symmetry group of $K_3$ are more 
convenient (see e.g. \cite{Kiritsis:2000zi}).

Irrespective of the choice of coset representative, the invariant metric 
on the second factor in \eqref{m4} can be obtained by 
decomposing  the right-invariant one-form $\theta_{6,n_v}=
de_{6,n_v}\cdot e_{6,n_v}^{-1}$ into a sum $h_{6,n_v} + p_{6,n_v}$
of  its compact and non-compact parts, and forming a quadratic combination
of the non-compact part $p_{6,n_v}$ which is invariant
under the action of the maximal compact subgroup 
$SO(6)\times SO(n_v)$. Combining it with the standard line element 
on the upper-half plane, the moduli space metric is thus given by 
 \be
\label{dsm40}
ds^2_{\cM_4}= \frac{d\tau_1^2+d\tau_2^2}{\tau_2^2} + {\rm Tr}(p_{6,n_v}^2)
= \frac{d\tau_1^2+d\tau_2^2}{\tau_2^2} 
- \frac14 {\rm Tr} (dM\cdot dM^{-1})
\ee
where $M\equiv e^T_{6,n_v}\cdot e_{6,n_v}$ is a symmetric matrix in 
$SO(6,n_v,\IR)$, invariant. Under the action of an element $g\in SO(6,n_v)$,
$e_{6,n_v}$ transforms by right-multiplication by $g$ followed by a 
compensating left-multiplication by an element in $SO(6)\times SO(n_v)$
so as to restore the gauge choice \eqref{e6nviwa} , while
$M$ transforms linearly in the symmetric representation $M\to g^T M g$.
 
Including the  $n_v+6$-dimensional gauge fields $A_{\mu\nu}^\Lambda$ 
($\Lambda=1\dots n_v+6$), arranged as a vector of $SO(6,n_v)$,
the complete bosonic action of $\cN=4$ supergravity at two-derivative
level is given by
\be
\begin{split}
S_4 = \int d^4 x \sqrt{-g} \left[  R \right.& - \frac12 \frac{d\tau_1^2+d\tau_2^2}{\tau_2^2}
+  \frac18 {\rm Tr} (dM\cdot dM^{-1}) \\
&\left. - \frac14 \tau_2  
F_{\mu\nu}^T \cdot M^{-1} \cdot F^{\mu\nu} 
+ \frac14 {\tau_1}F_{\mu\nu}^T \cdot \eta_{6,n_v} \cdot 
\tilde F^{\mu\nu} \right] \ .
\end{split}
\ee
While the action is manifestly invariant under $SO(6,n_v,\IR)$, the $Sl(2,\IR)$ symmetry 
is only visible at the level of the equations of motion.  According to string duality conjectures, 
the quantum theory is invariant under an arithmetic subgroup of 
$Sl(2,\IR)\times SO(6,n_v,\IR)$, whose precise definition depends on 
the model under consideration.

\subsection{Reduction to 3D\label{3d}}
In order to study stationary solutions, with metric 
\be 
\label{4d3d}
ds_4^2 = -e^{2U} (dt + \omega)^2 + e^{-2U} ds^2_3 \ ,
\ee
it is convenient to reduce the 4D $\cN=4$ supergravity theory along the time
direction to a $\cN=8$ theory supergravity in three Euclidean 
dimensions~\cite{Gunaydin:2007bg,Breitenlohner:1987dg,Sen:1994eb,
Cvetic:1995kv,Gunaydin:2005mx,Pioline:2006ni}.
After dualizing one-forms into pseudo-scalars, all bosonic
degrees of freedom can be described by a non-linear sigma model with 
non-Riemannian target space 
\be 
\label{m3star}
{\cal M}_3^*=G_3 / K_3^* = \frac{SO(8,n_v+2,\IR)}{SO(6,2)\times SO(2,n_v)}\ ,
\ee 
coupled to 3D Euclidean gravity. 
The moduli space \eqref{m3star} is related to the more familiar Riemannian
space arising in the reduction along a space-like 
direction \cite{Marcus:1983hb,Sen:1994wr}
\be 
\label{m3}
{\cal M}_3=G_3 / K_3 = \frac{SO(8,n_v+2,\IR)}{SO(8)\times SO(n_v+2)}
\ee 
by analytic continuation, as we describe presently. As in \cite{Obers:2000ta}, 
it is convenient to parametrize ${\cal M}_3$ by choosing a metric
\be
\label{eta262}
\eta_{8,n_v+2}=\begin{pmatrix}&&\CI_2\\&\eta_{6,n_v}&\\\CI_2&&\end{pmatrix}
\ee
and a coset representative in (partial) Iwasawa gauge,
\be
\label{iwa}
e_{8,n_v+2}=\begin{pmatrix}
\begin{array}{cc}\frac{e^{-U}}{\sqrt{\tau_2}}
&\frac{e^{-U}\tau_1}{\sqrt{\tau_2}}\\0&e^{-U}\sqrt{\tau_2}
\end{array}&0&0\\0&e_{6,n_v}&0\\0&0&
\begin{array}{cc}e^U\sqrt{\tau_2}&0\\-\frac{e^U \tau_1}{\sqrt{\tau_2}}
&\frac{e^U}{\sqrt{\tau_2}}\end{array}
\end{pmatrix}
\cdot
\begin{pmatrix}1&0 & \zeta^\Lambda  &
-\frac{1}{2}\zeta^\Lambda \zeta_\Lambda&\sigma-\frac{1}{2}\zeta^\Lambda \tilde{\zeta}_\Lambda \\
0 & 1 &  \tilde{\zeta}_\Lambda  &
-\sigma-\frac{1}{2}\zeta^\Lambda \tilde{\zeta}_\Lambda  &-\frac{1}{2}\tilde{\zeta}_\Lambda \tilde{\zeta}^\Lambda \\
0&0&\CI_{6,n_v}&-\zeta_\Lambda &-\tilde{\zeta}^\Lambda \\
0&0&0&1&0 \\
0&0&0&0&1\end{pmatrix}
\ee
with $e_{6,n_v}\in SO(6,n_v)/SO(6)\times SO(n_v)$ as in
\eqref{e6nviwa} or \eqref{e6nvas}. The coordinates $\zeta^\Lambda$
and $\tzeta_\Lambda$
correspond to the time-like component
of the gauge fields $A_{\mu\nu}^\Lambda $ and their magnetic dual, while 
$\sigma$ is the pseudo-scalar dual to the one-form $\omega$. 
The indices on $\zeta^\Lambda$
and $\tzeta_\Lambda$ are raised and lowered using the metric $\eta_{6,n_v}$.
The decomposition \eqref{iwa} reflects the fact that under the subgroup
$\IR^+\times Sl(2)\times SO(6,n_v)$,  $SO(8,n_v+2)$ admits the 
"real"  5-grading
\be
\label{real5g}
1\vert_{-2} \oplus (2,n_v+6)\vert_{-1} \oplus 
[ (1,1)\oplus (3,1)\oplus +(1,so(n_v+6)]\vert_0 \oplus 
(2,n_v+6)\vert_{1} \oplus 1\vert_{2} 
\ee
where the subscript indicates the charge under the $\IR^+$ factor generated by the 
diagonal matrix $(\CI_2,0_6,-\CI_2)$. The adjective "real" refers to the fact that each summand 
is invariant under the Cartan involution, so that the corresponding coordinates are real. 

The invariant
metric on \eqref{m3} is obtained by the same prescription as above \eqref{dsm40}, 
namely by decomposing the right-invariant one-form
\be
\theta_{8,n_v+2}=de_{8,n_v+2}\cdot e_{8,n_v+2}^{-1} 
\ee
into its compact  and a non-compact parts, and taking the
$SO(8)\times SO(n_v+2)$ invariant norm of the non-compact part. This is most easily
done by changing basis such that the maximal compact subgroup corresponds to 
square blocks of size $8$ and $n_v+2$ on the diagonal\footnote{The reason for 
choosing an off-diagonal metric for the $SO(8)$ part will become apparent shortly.},
\be
\label{eta82}
\eta_{8,n_v+2,K}=
\begin{pmatrix} & \CI_4 &  \\ \CI_4 & & \\ & & -\CI_{n_v+2} \end{pmatrix} = 
\Omega_K^T \,\eta_{8,n_v+2} \,\Omega_K 
\ee
Such a change of basis is non-unique; a convenient choice is
\footnote{This choice ensures that
the $SU(3)$ subgroup of the 4D R-symmetry group $SO(6)$ is mapped to a subgroup
$\scriptsize\begin{pmatrix} 1 & & & \\ & * &*&* \\  & * &*&* \\ & * &*&* \end{pmatrix}$
inside $SU(4)\subset SO(8)$.}
\be
\Omega_K= {\scriptsize
\left(
\begin{array}{lllllllllll}
 \frac{1}{2} & 0 & 0 & 0 & \frac{1}{2} & 0 & 0 & 0 & 0 &
   \frac{1}{\sqrt{2}} & 0 \\
 -\frac{i}{2} & 0 & 0 & 0 & \frac{i}{2} & 0 & 0 & 0 & 0 & 0 &
   \frac{1}{\sqrt{2}} \\
 0 & \frac{1}{\sqrt{2}} & 0 & 0 & 0 & \frac{1}{\sqrt{2}} & 0 & 0 & 0 & 0
   & 0 \\
 0 & 0 & \frac{1}{\sqrt{2}} & 0 & 0 & 0 & \frac{1}{\sqrt{2}} & 0 & 0 & 0
   & 0 \\
 0 & 0 & 0 & \frac{1}{\sqrt{2}} & 0 & 0 & 0 & \frac{1}{\sqrt{2}} & 0 & 0
   & 0 \\
 0 & -\frac{i}{\sqrt{2}} & 0 & 0 & 0 & \frac{i}{\sqrt{2}} & 0 & 0 & 0 & 0
   & 0 \\
 0 & 0 & -\frac{i}{\sqrt{2}} & 0 & 0 & 0 & \frac{i}{\sqrt{2}} & 0 & 0 & 0
   & 0 \\
 0 & 0 & 0 & -\frac{i}{\sqrt{2}} & 0 & 0 & 0 & \frac{i}{\sqrt{2}} & 0 & 0
   & 0 \\
 0 & 0 & 0 & 0 & 0 & 0 & 0 & 0 & \CI_{n_v} & 0 & 0 \\
 \frac{1}{2} & 0 & 0 & 0 & \frac{1}{2} & 0 & 0 & 0 & 0 &
   -\frac{1}{\sqrt{2}} & 0 \\
 -\frac{i}{2} & 0 & 0 & 0 & \frac{i}{2} & 0 & 0 & 0 & 0 & 0 &
   -\frac{1}{\sqrt{2}}
\end{array}
\right)}
\ee
In this new basis, the Cartan decomposition of $\theta_{8,n_v+2}$
is just the decomposition into blocks of dimension $8\times 8$,
$(n_v+2)\times (n_v+2)$ and $8 \times (n_v+2)$:
\be 
\label{thpbase}
\Omega_K^{-1}\, \theta_{8,n_v+2}\,\Omega_K = 
\begin{pmatrix} \theta_{AB} & p_{Aa} \\  p_{Aa} & \theta_{ab} \end{pmatrix}
\ee
where $A,B=1\dots 8, a=1\dots n_v+2$. Conventionally, we take the non-compact 
part $p_{Aa}$ to transform as a spinor of positive chirality under $SO(8)$. 
The quadratic form $\scriptsize \begin{pmatrix} 0 & \CI_4 \\\CI_4 & 0 \end{pmatrix}$
appearing in \eqref{eta82} is recognized as the charge conjugation matrix
$C_{AB}$ in the spinor representation (see Appendix \ref{so8gamma}
for our conventions for $SO(8)$ spinors). The compact parts 
$ \theta_{AB}$ and $ \theta_{ab}$ correspond to the 
$SO(8)$ and $SO(n_v+2)$ spin connections, respectively. Thus,
the right-invariant metric on $\cM_3$ is given by
\be
\label{dsm3}
ds^2_{{\cal M}_3}
=  p^{Aa} p^{Bb} C_{AB} \delta_{ab} \equiv g_{mn} d\phi^m d\phi^n \ .
\ee
The final result is 
\be
\label{dsm30}
ds^2_{{\cal M}_3}=
dU^2+ds^2_{{\cal M}_4}+
e^{-2U}
\frac{(d\zeta^\Lambda +\tau d\tilde\zeta^\Lambda)\cdot M\cdot 
(d\zeta^\Lambda +\bar\tau d\tilde\zeta^\Lambda)}
{\tau_2}+e^{-4U}(d\sigma+\zeta^\Lambda 
d\tilde{\zeta}_\Lambda -\tilde{\zeta}_\Lambda d\zeta^\Lambda )^2
\ee
where $\tzeta_\Lambda,\tzeta^\Lambda, \sigma$ are identified as the 
time-component of the gauge field $A_\Lambda$ and its magnetic 
dual $\tilde A^\Lambda$, and the NUT scalar dual dual to the
connection one-form $\omega$ in \eqref{4d3d}. 
This relation between the moduli spaces in 3D and 4D is 
a straightforward generalization of the $c$-map encountered in the
dimensional reduction of $\cN=2$ theories \cite{Ferrara:1989ik}. 
As mentioned above, the indefinite metric on the manifold ${\cal
M}_3^*$ is obtained from \eqref{dsm3} by analytically continuing
$(\zeta^\Lambda ,\tilde{\zeta}_\Lambda )\rightarrow -i(\zeta^\Lambda ,\tilde{\zeta}_\Lambda )$.

The supersymmetrization of the non-linear sigma model on $\cM_3$ was studied in 
detail in  \cite{Marcus:1983hb,de Wit:1992up,deWit:2004yr}. We briefly summarize
the main results following \cite{deWit:2004yr}. 
The $\cN=8$ supersymmetry algebra relies on the existence of seven almost complex 
hermitian structures $f^{P m}_n$ ($P=2\dots 8$) satisfying the $SO(7)$ Clifford algebra. 
From these, one may construct 28 two-forms $f^{\mu\nu}=f^{\mu\nu}_{mn}
d\phi^m d\phi^n$ ($\mu=1\dots 8$) via
\be
f^{PQ}_{mn}=f^{[P p}_m f^{Q] q}_n \, g_{pq}\ ,\quad 
f^{1P}_{mn}=-f^{P1}_{mn}= g_{mp} f^{P p}_n \ .
\ee
The tensors $f^{\mu\nu}$ are covariantly constant, and equal to the 
curvature of the $SO(8)$ spin connection $Q^{\mu\nu}=Q^{\mu\nu}_{m}d\phi^m$,
\be
dQ^{\mu\nu} + 2 Q^{\rho[\mu} \wedge Q^{\nu]\rho} = \frac12  f^{\mu\nu}\ .
\ee 
The fermionic degrees of freedom are most easily described by 
introducing a fermionic tensor $\chi^{m\mu}$ subject to the constraint
\be
\label{conferm}
\chi^{m\mu} = \frac18 \left( \delta^{\mu}_{\nu} \delta^m_n - f^{\mu m}_{\nu n} \right) \chi^{n\nu} \ ,
\ee
which projects down the number of components to $8(n_v+2)$. The 
supersymmetry variations of the gravitini $\psi^\mu_M$ ($M=1,2,3$) 
and the dilatini  $\chi^{m\mu}$, for vanishing fermionic background,  
are then written as
\bea
\delta \psi^\mu_M = D_M\epsilon^{\mu} \ ,\quad
\delta \chi^{m\mu} = \frac12 ( \delta^{\mu\nu} \delta^{mn} -f^{\mu\nu}_{mn} ) 
\gamma^M \pa_M \phi^n \eps^\nu 
\eea
For our purposes, it will be convenient to solve the constraint \eqref{conferm}
explicitly, as
\be
\chi^{m\mu} = e^m_{A a} \Gamma^{\mu}_{AA'} \lambda^{a A'}
\ee
where $e^m_{A a}= (p^{A a}_m)^{-1}$ is the inverse viel-bein afforded by
the $SO(8)\times SO(n_v+2)$ restricted holonomy, and $\Gamma^{\mu}_{AA'}$
are the $SO(8)$ sigma matrices. In terms of the unconstrained spinor $\lambda^{a A'}$,
the variation of the dilatini is given by 
\be
\delta \lambda^a_{A'} = p^{A a}\,\Gamma^{\mu}_{A'A}\, \vareps_\mu\ .
\ee
Notice that the supersymmetry parameter $\eps^\mu$, dilatini 
$\lambda^{a A'}$ and bosonic derivatives $p^{A a}$ transform as the
three inequivalent 8-dimensional representations of the R-symmetry
group. Of course, one could use triality and permute the
representations assigned to these objects.

\subsection{Reduction to 1D\label{1d}}
Upon further restricting to spherically symmetric solutions, with spatial 
metric
\be
ds_3^2 = N^2(\rho)d\rho^2+r^2(\rho)(d\theta^2+\sin^2\theta\, d\phi^2)\ ,
\ee
the 3D non-linear sigma model reduces to the geodesic motion of a free particle
on a real cone $\IR^+\times \cM_3^*$ over \eqref{m3star}, with action
\be
\label{L1}
S_1=\int d\rho \left[ \frac{N}{2} + \frac{1}{2N} \left( r'^2 - r^2 
g_{mn} \phi'^m \phi'^n \right) \right]\ .
\ee
The equation of motion of $N$ forces the Hamiltonian to vanish,
\be
\label{wdw}
H_{\rm WDW} = (p_r^2) - \frac{1}{r^2} g^{mn} p_{\phi^m} p_{\phi^n} -1 \equiv 0
\ee
The system reduces to geodesic motion on $\cM_3^*$, with momentum
squared $g^{mn} p_{\phi^m} p_{\phi^n}=C^2$, and motion along $r$
with conformally invariant Hamiltonian $(p_r^2)-C^2/r^2-1=0$. In
particular, the phase space is given by the symplectic quotient of the 
cotangent bundle $T^*(\IR^+ \times \cM_3^*)$ by the first class
constraint $H_{\rm WDW}=0$. Extremal
black holes necessarily have $C^2=0$ (although this condition is not 
sufficient), which gives a further first class constraint.

By the usual Noether procedure, Killing vectors $\kappa^m \pa_{\phi^m}$ 
of $\cM_3^*$ yield conserved  quantities $ \kappa^m p_{\phi^m}$ for 
the geodesic motion  on $\cM_3^*$. Of particular interest are
the isometries corresponding to shifts in the $\zeta,\tzeta,\sigma$ directions,
\be
\label{pqk}
P^\Lambda = \pa_{\tzeta_\Lambda} -\zeta^\Lambda \pa_\sigma\ ,\quad
Q_\Lambda = -\pa_{\zeta^\Lambda} -\tzeta_\Lambda \pa_\sigma\ ,\quad
K = \pa_\sigma
\ee
which satisfy the Heisenberg algebra
\be
\label{heisr}
\left[ P^\Lambda , Q_\Sigma\right] = -2 \delta^\Lambda_\Sigma K\ . 
\ee
Bona fide black holes have zero NUT charge $K=0$, 
in which case $P^\Lambda , Q_\Sigma$ correspond
to the electric and magnetic charges of the black hole. In addition, the conserved
quantity associated to the Killing vector
\be
H= - \pa_U - \zeta^\Lambda \pa_{\zeta_\Lambda}- \tzeta_\Lambda \pa_{\tzeta_\Lambda}\\
\ee
is the ADM mass, provided one enforces the condition
\be
U=\zeta^\Lambda =\tilde{\zeta}_\Lambda =\sigma=0
\ee
at spatial infinity.

While the conserved charges $P^\Lambda, Q_\Lambda, K, H$ appear  universally in reductions 
of Einstein-Maxwell theories, in the present case there are additional conserved quantities 
due to the  isometries of the scalar moduli space in 4 dimensions. For $n_v=0$,
the corresponding Killing vectors read 
\bea
\label{ysl2}
Y_0 &=& \tau_1 \pa_{\tau_1} + \tau_2 \pa_{\tau_2}+ \frac12 
\zeta^\Lambda \pa_{\zeta_\Lambda}- \frac12 \tzeta_\Lambda \pa_{\tzeta_\Lambda} \\
Y_+ &=& \pa_{\tau_1} - \tzeta_\Lambda \pa_{\zeta^\Lambda}\\
Y_- &=& \frac12 (\tau_1^2-\tau_2^2)  \pa_{\tau_1} + \tau_1 \tau_2 \pa_{\tau_2} + \frac12 
 \zeta^\Lambda \pa_{\tzeta_\Lambda}\
\eea
and satisfy the $Sl(2,\IR)$ commutation relations
\be
\left[ Y_0, Y_\pm\right] = \pm Y_\pm \ ,\quad \left[ Y_+, Y_-\right] = Y_0 \ . 
\ee

In addition to the bosonic terms displayed in \eqref{L1}, the one-dimensional 
Lagrangian contains fermionic terms corresponding to the reduction of 
the $\cN=8$ supersymmetric sigma model in 3 dimensions along the sphere.
This reduction was studied in detail in \cite{Gunaydin:2007bg}
in the $\cN=4$ case, and it was found that the reduction yields a 
one-dimensional sigma model with the same number of (spinorial)
supersymmetries as in 3 dimensions \footnote{To be precise, the 
supersymmetric completion of the 1D sigma model is known only
in the sector involving $\cM_3$ but not $r$ and $N$.}. Following the
same analysis, we find that the conditions for radially symmetric 
solutions to preserve supersymmetry are given by
\be
\label{susycond}
\exists \vareps_{\mu}\in\IC^8\backslash\{0\} \ \slash \ \forall\, a,A'\ ,\quad 
p^{A a}\,\Gamma^{\mu}_{A'A}\, \vareps_\mu=0
\quad \mbox{and} \quad r'=N\ .
\ee
The first condition implies that any linear combination of the $n_v+2$ 
spinors $p^{Aa}$ has zero norm. Put differently, 
\be
\label{p2}
p^{A a} p^{B b} C_{AB} = 0 \ .
\ee
This condition is in fact equivalent to the existence of  $\vareps_{\mu}$
such that \eqref{susycond} is obeyed\footnote{By an $SO(8)$ rotation, the first
spinor $p^{A1}$ (rotated by the $8\times 8$ upper-left block of $\Omega_K$)
can be chosen parallel to $(1,0,0,0,i,0,0,0)$; the second can be chosen to lie
along $(0,1,0,0,0,i,0,0)$ up to the addition of the first, etc. In this basis, it is
easy to check that all $p^{Aa}$ are annihilated by $\Gamma^{\mu}_{A'A}\, \vareps_\mu$
with $\vareps_\mu=\sqrt{2}/2(1,0,0,0,0,0,0,-i)$, corresponding to $Y_i=0$
in \eqref{epsy} below.}. Clearly, it implies the extremality condition $C^2=
p^{A a} p^{B b} C_{AB} \delta_{ab}=0$, but is considerably stronger.
In Section 3.4, we shall explain how it can be expressed  as holomorphic
geodesic motion on the twistor space $Z$. For what concerns
the second condition $r'=N$, it is consistent with the condition 
$p_r=\pm 1$ following from the Hamiltonian constraint
\eqref{wdw} at extremality,  but implies that  only the choice
of the upper sign in this relation is consistent with supersymmetry.

\subsection{Geodesics and one-parameter subgroups}
Since the target space  $\cM_3^*$ is a symmetric space, all geodesics
correspond to one-parameter subgroups in $G_3$.
A geodesic passing through the point $e_0$ at $\tau=0$ with initial velocity
$p_0$ is given by 
\be
e(\tau)=k(\tau)\cdot e^{p_0 \tau/2}\cdot e_0\ ,
\quad
M(\tau)\equiv e^T(\tau)\cdot e(\tau) = e_0^T\cdot e^{p_0 \tau} \cdot e_0
\ee
where $p_0$ is a non-compact (i.e. symmetric) element in $\mathfrak{g}_3$, 
$k(\tau)$ is the unique element of $K_3$ which brings $e(\tau)$ back
to the Iwasawa gauge, and $\tau$ is the affine parameter. 
The $\mathfrak{g}_3$-valued conserved charge inherited 
from the right action of $G_3$ is then given by
\be
Q = -dM\,M^{-1} = - e_0^{t}\, p_0\, e_0^{-t}\ .
\ee
The velocity $p_0$ may be traded for the Noether charge $Q$, but it
should be noted that the latter cannot be chosen independently from 
the initial position $M_0$, since $Q M= M Q^T$ at all times. In 
terms of $Q$, the geodesic motion is given by 
\be
\label{solgeo}
e(\tau)=k(\tau)\cdot e_0 \cdot e^{-Q^T \tau/2}\ ,\quad
M(\tau) = e^{-Q \tau/2}\cdot M_0 \cdot e^{-Q^T \tau/2}
\ee
The affine parameter $\tau$ is equal to the radial parameter 
$\rho$ in the gauge $N(\rho)=r^2(\rho)$. The motion of $r(\rho)$ 
may be obtained by integrating the Hamiltonian
constraint \eqref{wdw}, and depends only on $p_0^2$.

The action of an element $g$ of $G_3$ takes the solution
\eqref{solgeo} to another solution with $Q\to g^T Q g^{-T},
M_0\to g^T M_0 g$. As a result, trajectories may be classified
according to the  orbit of the matrix of Noether charges
$Q$ under the co-adjoint action of $G_3$. Of special interest are nilpotent orbits,
i.e. those for which $Q^r=0, Q^{r-1}\neq 0$ for some $r\geq 2$
(the degree $r$ depends on representation in which $Q$ is 
evaluated; here we consider the defining representation of $G_3$).
Indeed, it was pointed out in \cite{Gunaydin:2005mx} that BPS
black holes in very special $\cN=2$ supergravity theories correspond 
to specific nilpotent orbits of degree 3. Subsequently, it was shown
that for very special $\cN=2$ supergravity with one vector multiplet, 
nilpotent orbits  of degree 3 yield (in general non-BPS) extremal black holes 
in 4 dimensions \cite{Gaiotto:2007ag}. It is straightforward to check that 
the argument in \cite{Gaiotto:2007ag}
extends to the present case. It is therefore interesting to determine 
the allowed nilpotent orbits of degree 3 for $G_3=SO(8,n_v+2)$.

Since $Q$ is conjugate to $p_0={\scriptsize\begin{pmatrix} 0 & p^{Aa} \\ 
p^{Aa} & 0\end{pmatrix}}$ in the basis \eqref{thpbase}, the condition $Q^3=0$ 
amounts to 
\be
\label{p3}
p^{A a} p^{B b} p^{Cc} \,C_{AB} \,\delta_{bc}= 0 \ .
\ee
This condition is clearly obeyed by BPS solutions, which satisfy the quadratic
constraint \eqref{p2}.
In fact, one may check explicitly that (for $n_v\geq 2$) the Noether charge for 
BPS solutions lies in the orbit $(3^4,1^{n_v-2})$ of the complexified
group $SO(10+n_v,\IC)$ (see Appendix B for 
a review of general facts about nilpotent orbits, and a Table of the low-dimensional 
nilpotent orbits of orthogonal groups). This follows from the fact, to be discussed in
Section 3, that BPS trajectories can be lifted to holomorphic geodesics on the
twistor space $Z$, which is equal to the orbit $(3^4,1^{n_v-2})$  via \eqref{comp5g2}.
This orbit is the "largest" nilpotent orbit of degree 3 (amongst orbits with dimension
less than $\mathcal{O}(8 n_v)$), in the sense that it intersects
the closure of any orbit of degree 3 (as apparent on Figure 1). 
This identification implies that the phase space of 1/4-BPS solutions 
in $\cN=4$ supergravity with $n_v$ vector multiplets is $8n_v+28$ 
dimensional\footnote{This is before enforcing the
first class constraint $K=0$.},
much larger than the dimension $4n_v+26$ of the phase space of 1/2-BPS solutions 
in a $\cN=2$ supergravity with the same number $n_v+6$ of 
vector fields \cite{Gunaydin:2007bg}. The 
extra degrees of freedom correspond to the $n_v$ hypermultiplets  coming from
the decomposition of the $n_v$ $\cN=4$ vector multiplets. The twistor techniques of
the next section in principle allow to find the most general 1/4-BPS solution, although
we fall short of this goal due to technical difficulties explained in Section 3.4. 
 
For what concerns non-BPS extremal black holes, they correspond to solutions of \eqref{p3}
which do not satisfy \eqref{p2}. Since there exist  (at least) two different real nilpotent 
orbits of type $(3^4,1^{n_v-2})$, related by an outer automorphism of $SO(8,n_v+2)$,
it is natural to conjecture that such a transformation will map BPS solutions
to non-BPS extremal solutions. Finding the general form of these non-BPS solutions
is outside the scope of this paper.

\section{Twistorial techniques for $\cN=4$ BPS black holes\label{twisec}}

We now return to the supersymmetry condition \eqref{susycond}, and introduce
geometric methods which allow to implement these constraints, both at a 
classical and quantum level in a convenient way. We work with the original 
Riemannian space \eqref{m3}, and perform analytic continuations at the end.

As emphasized in \cite{Gunaydin:2007bg}, it is expedient to eliminate the 
existence quantifier in \eqref{susycond} by enlarging the phase space
with the complex Killing spinor $\varepsilon_\mu $. Since the latter is always of
zero norm and defined up to the action of $\IC^\times$, it is best
viewed  as an element of the
complex symmetric space 
\be
F=\frac{SO(8,\IR)}{U(4)} \sim \frac{SO(8,\IR)}{SO(2)\times SO(6)}
\ee
As we explain in more detail below, this equality 
reflects the fact
that Cartan pure spinors in 8 dimensions are just zero norm spinors.
Remarkably,  it is possible to fiber\footnote{Note that unlike the 
quaternionic-K\"ahler case, the fiber is {\it not} the 
sphere of almost complex structures $S^6$, but a complexification thereof} 
$F$ over $\cM_3$ such that the total 
space $Z$ admits
an integrable complex structure \cite{Bryant}: this is achieved by
``cancelling the $SO(8)$ factors'', namely by considering the
homogeneous (but not symmetric) complex space $Z\equiv M_3\backslash 
G_3$
\be
\label{defZ}
Z = \frac{SO(8,n_v+2,\IR)}{U(1)\times SU(4)\times SO(n_v+2)}\sim
\frac{SO(8,n_v+2,\IR)}{SO(2)\times SO(6)\times SO(n_v+2)}\ .
\ee
The integrable complex structure is afforded by the $U(1)=SO(2)$ factor in the denominator. 
Moreover, as we show below, the
BPS conditions \eqref{susycond} guarantee that the geodesic motion
on $\cM_3$ can be lifted to a holomorphic curve on $Z$. 
This construction parallels the $\cN=2$ case discussed in \cite{Neitzke:2007ke}, 
upon replacing the complex projective twistor line $\IC P^1$ with the Grassmannian $F$.

\subsection{Parametrizing the fiber \label{secfib}}
In a basis where the invariant metric 
takes the off-diagonal block form $\eta_8= {\scriptsize 
\begin{pmatrix}&\CI_4\\\CI_4&\end{pmatrix}}$, a coset representative of $SO(8,\IR)$
may be chosen as 
\be
\label{so8u4}
e_F = \begin{pmatrix}\CI_4&0\\\bar{X}&\CI_4\end{pmatrix}\cdot
\begin{pmatrix}1/\sqrt{1-X\bar{X}}&0\\0&\sqrt{1-\bar{X}X}\end{pmatrix}
\cdot\begin{pmatrix}\CI_4&X\\0&\CI_4\end{pmatrix}\ .
\ee
where $X_{IJ}$ ($I,J=1\dots 4$) is a $4\times 4$ 
antisymmetric complex\footnote{Since we are 
dealing with
the compact form of $SO(8)$, the matrix representation in this basis has 
to be complex. The split form $SO(4,4)$ would instead be obtained by taking
$X$ and $\bar X$ as independent real variables.} matrix $X$. 
This decomposition realizes the Harish-Chandra
embedding $K\backslash G(\IR) \hookrightarrow P(\IC)\backslash G(\IC)$
where $G(\IR)=SO(8,\IR)$ and $P(\IC)$ is the parabolic subgroup
of lower block-triangular matrices of the form 
$\scriptsize\begin{pmatrix}*&\\ *&*\end{pmatrix}$,
and guarantees that $X$ are complex coordinates on $F$.
Moreover, it makes explicit the holomorphic action of $G({\IC})$
on $F$, by right multiplication on \eqref{so6so2HC} 
followed by left multiplication by an element of $P({\IC})$.
On general grounds~\cite{Satake}, a K\"ahler potential for the 
invariant metric on $F$ is given by the logarithm of a character
of $K(\IC)=GL(4,\IC)$ evaluated on the block-diagonal component
in the Harish-Chandra decomposition \eqref{so8u4},
\be 
\label{KX}
K(X,\bar X)=\log \det(1-X\bar{X})\ .
\ee
The first four rows of the  right-most matrix in \eqref{so8u4}
define an isotropic\footnote{i.e. a 4-plane of zero norm vectors:
$(\CI_4 \,|\,X)^T\eta_8 (\CI_4 \,|\,X)=0$ since $X=-X^T$.}
4-plane  $\IC^4=(\CI_4 \,|\,X)$ inside $\IC^8$. Such
isotropic planes are also known as projectivized pure
spinors in Cartan's sense.

On the other hand, in a basis where the invariant metric takes the
form $\tilde\eta_8=\scriptsize\begin{pmatrix}&&1\\ &\CI_6& \\ 1&&\end{pmatrix}$,
a coset representative of $SO(8,\IR)/SO(2)\times SO(6)$ may be chosen as 
\be
\label{so6so2HC} 
\tilde e_F=
\begin{pmatrix}1&&\\\bar{Y}^i&\CI_6&\\-\frac{1}{2}\sum\bar{Y}_k^2&-\bar{Y}_i&1
\end{pmatrix}\cdot
\begin{pmatrix}e^{-K(Y,\bar Y)/2}&&\\&[A(Y,\bar Y)]^{-1/2}&\\
&&e^{K(Y,\bar Y)/2}
\end{pmatrix}\cdot\begin{pmatrix}1&Y_i&-\frac{1}{2}\sum Y_k^2\\&\CI_6&-Y_i\\&&1
\end{pmatrix}
\ee 
where  the scalar $e^{K(Y,\bar Y)}$ and the $6\times 6$ matrix
$A(Y,\bar Y)$ are determined in terms of the complex
coordinates $Y_i$ and their complex conjugate $\bar Y_i$:
\begin{equation}
\label{KY}
K(Y,\bar Y) = 
\frac{1}{2}\log\left(1+\sum_i Y_i \bar Y_i+\frac{1}{4} 
(\sum_i Y_i^2) (\sum_i \bar Y_i^2) \right)\ ,
\end{equation}
\be
A_{ij}(Y,\bar Y) = \delta_{ij}+ e^{K} \left[ Y_i \bar Y_j -  Y_j \bar Y_i 
- \frac12 Y_i Y_j  (\sum \bar Y_k^2)- \frac12  \bar Y_i \bar Y_j (\sum Y_k^2)
+ Y_i \bar Y_j (\sum Y_k \bar Y_k) \right]\ .
\ee
Again, $K(Y,\bar Y)$ provides the K\"ahler potential for the $SO(8)$-invariant 
K\"ahler metric on $F$. This time, the first row $( 1 , Y_i, -
\frac12 \sum Y_k^2)$ in the right-most matrix
in \eqref{so6so2HC} provides the most general null vector for the metric 
$\eta_8$, up to a $\IC^\times$ action. Thus, in eight dimensions Cartan pure 
spinors are indeed the same as projectivized null vectors.

Based on this observation, it is natural to identify this null vector 
with the Killing spinor $\vareps_\mu$,
\be
\label{epsy}
\vareps_\mu = \left( \frac{2- \sum Y_k^2}{2\sqrt2}  , Y_i,  \frac{2+ \sum Y_k^2}{2i\sqrt2} \right)\ ,\quad
\sum_\mu \vareps_\mu^2 = 0 \ .
\ee
To see the relation to the coordinates $X_{IJ}$  note that for 
a fixed null vector $\vareps_\mu$, the equations
\be
\label{ga}
\forall A'\ ,  \quad \vareps_\mu  \,\Gamma^\mu_{A' A} \, p^{A}  = 0 
\ee
select an isotropic 4-plane in the 8-dimensional space of the spinors $p^{A}$. 
In fact, using the explicit representation of the $SO(8)$ Dirac matrices 
given in Appendix \ref{so8gamma}, we have the rank 4 matrix
\be
\vareps_\mu  \,\Gamma^\mu_{A' A}  = 
\begin{pmatrix}
i \sqrt2 \,\CI_4 & \sum_k Y_k \Sigma^k \\  \sum_k \tilde Y_k \Sigma^i & -\frac{i}{\sqrt2}
\sum_k Y_k^2 
\end{pmatrix}
\ee 
where $\Sigma_i$ are $SO(6)$ Sigma matrices.
Identifying the first four rows of this matrix with the isotropic 
4-plane $(\CI_4 \,| \,X)_{I A}$ leads to the relation between the $X$ and $Y$
coordinates,
\be
\label{XtoY}
X_{IJ}= -\frac{i}{\sqrt2} Y_i \, \Sigma^i_{IJ} \ .
\ee
It may be checked explicitly that the K\"ahler potentials
\eqref{KY} and \eqref{KX} agree, up to a K\"ahler transformation.

\subsection{Bottom-up construction of the twistor space}

The homogeneous complex space $Z$ defined in \eqref{defZ} may be parameterized by
relaxing the Iwasawa gauge in  \eqref{iwa}, and introducing a coset representative
$\hat e_F$  of the fiber $U(4)\backslash SO(8)$,
\be
\label{Zef}
e_Z = \hat{e}_F \cdot e_{8,n_v+2} \ , \quad 
\ee
where $\hat{e}_F$ is obtained by embedding $e_F$ inside
the maximal compact subgroup $SO(8)\times SO(n_v+2)$ of $G_3$,  
\be
\label{hatef}
\hat{e}_F = \Omega\cdot 
 \begin{pmatrix}\CI_4&0&0\\ 0&\CI_{n_v+2} & 0 \\
 \bar{X}&0 & \CI_4\end{pmatrix}\cdot
\begin{pmatrix}1/\sqrt{1-X\bar{X}}&0&0\\
0&\CI_{n_v+2} & 0 \\
0&0&\sqrt{1-\bar{X}X}\end{pmatrix}
\cdot\begin{pmatrix}\CI_4&0&X\\0&\CI_{n_v+2} & 0 \\
0&0&\CI_4\end{pmatrix}  \cdot \Omega^{-1}\  .
\ee
Here, the matrix 
\be
\label{omc}
\Omega = {\scriptsize
\left(
\begin{array}{lllllllllll}
 \frac{1}{2} & 0 & 0 & 0 & 0 & \frac{1}{\sqrt{2}} & 0 & \frac{1}{2} & 0 &
   0 & 0 \\
 -\frac{i}{2} & 0 & 0 & 0 & 0 & 0 & \frac{1}{\sqrt{2}} &
   \frac{i}{2} & 0 & 0 & 0 \\
 0 & \frac{1}{\sqrt{2}} & 0 & 0 & 0 & 0 & 0 & 0 & \frac{1}{\sqrt{2}} & 0
   & 0 \\
 0 & 0 & \frac{1}{\sqrt{2}} & 0 & 0 & 0 & 0 & 0 & 0 & \frac{1}{\sqrt{2}}
   & 0 \\
 0 & 0 & 0 & \frac{1}{\sqrt{2}} & 0 & 0 & 0 & 0 & 0 & 0 &
   \frac{1}{\sqrt{2}} \\
 0 & -\frac{i}{\sqrt{2}} & 0 & 0 & 0 & 0 & 0 & 0 & \frac{i}{\sqrt{2}} & 0
   & 0 \\
 0 & 0 & -\frac{i}{\sqrt{2}} & 0 & 0 & 0 & 0 & 0 & 0 & \frac{i}{\sqrt{2}}
   & 0 \\
 0 & 0 & 0 & -\frac{i}{\sqrt{2}} & 0 & 0 & 0 & 0 & 0 & 0 &
   \frac{i}{\sqrt{2}} \\
 0 & 0 & 0 & 0 & \sqrt{2}\, \CI_{n_v} & 0 & 0 & 0 & 0 & 0 & 0 \\
 \frac{1}{2} & 0 & 0 & 0 & 0 & -\frac{1}{\sqrt{2}} & 0 & \frac{1}{2} & 0
   & 0 & 0 \\
 -\frac{i}{2} & 0 & 0 & 0 & 0 & 0 & -\frac{1}{\sqrt{2}} &
   \frac{i}{2} & 0 & 0 & 0
\end{array}
\right)}
\ee
provides the change of basis from the metric \eqref{eta262} to the
metric
\be
\label{eta424}
\hat\eta_{8,n_v+2}=
\begin{pmatrix} & & \CI_4 \\ & -\CI_{n_v+2} & \\ \CI_4 & & \end{pmatrix}
=\Omega^T \,\eta_{8,n_v+2} \,\Omega 
\ee
Block-diagonal matrices of the form 
\be
\begin{pmatrix}A&0&0\\0&B&0\\0&0&A^{-T}
\end{pmatrix}\ ,\quad A\in U(4), \ ,B\in SO(n_v+2)
\ee 
generate a $U(1)\times SU(4)\times SO(n_v+2)$ subgroup of the 
maximal compact subgroup $SO(8)\times SO(n_v+2)$. It is important to 
note that the $U(4)$ factor inside $SO(8)$ is distinct 
from the $U(1)\times SO(6)$ 4-dimensional R-symmetry group. These two groups
only share a $U(3)$ common subgroup, which is manifest with our choice
of $\Omega$ in \eqref{omc}.

Now, consider the decomposition of the Lie algebra of $SO(8,n_v+2)$ under $U(1)\times SU(4)\times 
SO(n_v+2)$, 
\be
\label{comp5g}
\bar 6\vert_{-2} \oplus (\bar 4,n_v+2)\vert_{-1} \oplus \left[  SU(4) \oplus U(1) \oplus SO(n_v+2) \right]_0
\oplus (4,n_v+2)\vert_{1} \oplus  6\vert_{2} 
\ee 
where the subscript indicates the $U(1)$ charge. Here, in contrast to the 
5-grading \eqref{real5g}, the Cartan involution exchanges the spaces of positive 
and negative charge. The right-invariant one-form
$\theta^Z = \Omega^{-1}\cdot de_Z e_Z^{-1}\cdot \Omega$ 
decomposes along each summand in \eqref{comp5g} as
\be
\theta^Z = \bar\theta_{IJ}^Z {\bar T}^{IJ}+  \bar\theta_{I a}^Z {\bar T}^{I a}+ 
(\theta_{SU(4)}^Z + \theta_{U(1)}^Z +\theta_{SO(n_v+2)}^Z) + 
\theta_{I a} ^Z{T}^{I a} + \theta_{IJ}^Z {T}^{IJ}
\ee
where the generators ${\bar T}^{IJ},{\bar T}^{I a},{T}^{I a} ,
{T}^{IJ}$ have charge $-2,-1,1,2$
respectively. On general grounds, the positive charge components $ \theta_{IJ},
\theta_{I a}^Z$ correspond to (1,0) forms on $Z$, while their complex conjugate 
$ \bar\theta_{IJ}^Z, {\bar \theta}^Z_{I a}$ are (0,1) forms.  In the basis corresponding to 
the metric \eqref{eta424}, the $U(1)$ factor is generated by the diagonal 
matrix ${\rm diag}(\CI_4,0_{n_v+2},-\CI_4)$, and therefore $\theta^Z_{IJ},
\theta_{I a}^Z$ are just the $4\times 4$ and $4 \times (n_v+2)$ blocks in the
upper triangular part of $\theta_Z$,
\be
\theta^Z = \begin{pmatrix} \theta_{SU(4)} ^Z+ \theta_{U(1)}^Z & 
\theta_{I a}^Z &  \theta_{IJ}^Z  \\
\bar\theta_{I a}^Z  & \theta_{SO(n_v+2)}^Z & \theta_{I a}^Z \\
\bar\theta_{IJ}^Z, &{\bar \theta}_{I a}^Z & -(\theta_{SU(4)}^Z)^\dagger- \theta^Z_{U(1)} 
\end{pmatrix}
\ee 
In terms of the components $p_{Aa}, \theta_{AB}$  of the right-invariant one-form
\eqref{thpbase} on the base $\cM_3$, the (1,0) forms read
\bea
\label{th10}
\theta_{I a}^Z &=& V ( \CI_4 \,| X )_{I A} \, p^{A a}\ ,\quad \\
\theta_{I J}^Z &=& V \left( dX + \theta^{(2)}  +  \theta^{(4)}  X - X \theta^{(1)}  - X \theta^{(3)}  X \right) V^T
\eea
where
\be
V = (1-X\bar{X})^{-1/2}
\ee
and $\theta^{(k)}$ are the $4\times 4$ blocks in the $SO(8)$ connection,
\be
\theta_{AB} = \begin{pmatrix} \theta^{(1)} _{IJ}&  \theta^{(2)}  _{IJ}\\  
\theta^{(3)}  _{IJ}&  \theta^{(4)}  _{IJ}
\end{pmatrix}\ .
\ee
Similarly, the $(0,1)$ invariant forms may be obtained from the lower triangular
part of $\theta_Z$, or by complex conjugation from the $(1,0)$ forms,
using the fact that  $\bar\theta^{(2)} = \theta^{(3)},  \bar\theta^{(1)}= \theta^{(4)}$:   
\bea
\label{th01}
\bar\theta^{I a}_Z &=& \bar V ( \bar X \,| \CI_4 )_{I A} \, p^{A a} ,\quad \\
\bar\theta^{I J}_Z &=& \bar V \left( 
d\bX + \theta^{(3)}  +  \theta^{(1)}  \bX - \bX \theta^{(4)}  - \bX \theta^{(2)}  \bX \right) \bar V^T
\equiv  d\bX + \overline{\mathcal{P}} 
\eea
Giving the (1,0) and (0,1) forms uniquely specify an almost complex
structure $\mathcal{J}$ on $Z$. Since linear combinations of (1,0) forms
stay of (1,0) type, we may set $V=1$ in \eqref{th10} and \eqref{th01} , 
and take as a basis of (1,0) forms
\bea
\label{th10s}
DX_{IJ} & \equiv& dX + \theta^{(2)}  +  \theta^{(4)}  X - X \theta^{(1)}  - X \theta^{(3)}  X 
\equiv dX + \mathcal{P} \ ,\quad \\
DZ_{I a}& \equiv&  ( \CI_4 \,| X )_{I A} \, p^{A a} \ .
\eea
In the next subsection, we shall show that $\mathcal{J}$ is in fact integrable. 
Observe that the (1,0)-forms $DZ_{I a}$ are linear combinations of the 
cotangent forms $p^{A a}$ whose coefficients are holomorphic functions
on the fiber, while the (1,0)-forms $DX$ are obtained by adding the
 "projectivized $SO(8)$ connection"  $\mathcal{P}$ to the 
 holomorphic differentials $dX$ on the fiber.  This directly parallels
 the twistor construction for \qk spaces.

 Finally, a family of invariant Hermitian metrics on $Z$ may be constructed by forming
 $SU(4)\times SO(n_v+2)$ invariant quadratic combinations of the (1,0) and (0,1)
 forms,
 \be
 \label{dsnu}
 ds^2 = \theta_{I J}^Z \bar\theta^{I J}_Z + \, \nu\, \theta_{I a}^Z \bar\theta^{I a}_Z\ .
 \ee
The parameter $\nu$ can be fixed by requiring that the metric is K\"ahler (see Section \ref{topdown}).
 
 \subsection{Top-down construction of the twistor space} \label{topdown}
We now describe an alternative construction of $Z$, which makes it manifest
that the almost complex structure $\mathcal{J}$ is integrable, and that $Z$
admits an invariant K\"ahler metric. As in our discussion of the K\"ahler
metric on the fiber in Section \ref{secfib}, and in analogy with \cite{Gunaydin:2007qq}, 
we rely on  the Harish-Chandra
embedding $M_3\backslash G_{3}(\IR) \hookrightarrow 
P_{3}(\IC)\backslash G_3(\IC)$ where  $P_{3}(\IC)$ is the parabolic subgroup
of lower block-triangular matrices in the basis where the metric takes the 
off-diagonal form 
\be
\eta_{8,n_v+2}=
\begin{pmatrix} & & \CI_4 \\ & -\CI_{n_v+2} & \\ \CI_4 & & \end{pmatrix}\ ,\quad
P_3(\IC) =\begin{pmatrix}*&&\\ *&*&\\ *&*&*\end{pmatrix}
\ee
This embedding is achieved by decomposing any element $g \in G_3$ as
a product
\be
\label{truehc}
g=\begin{pmatrix}*&&\\ *&*&\\ *&*&*\end{pmatrix}\cdot 
\begin{pmatrix}\CI_4&\mZ&
\mXt \\0&\CI_{n_v+2}&\mZ^T\\0&0&\CI_4
\end{pmatrix}
\ee
where $\mXt\equiv  \mX+\frac{1}{2}\mZ \mZ^T$ is an $4\times 4$ antisymmetric complex 
matrix 
and $\mZ$ is a $4\times (n_v+2)$ complex matrix. The map $g\in M_3(\IR)\backslash 
G_{3}(\IR) \hookrightarrow  (\mX,\mZ) \in P_{3}(\IC)\backslash G_3(\IC) $ is well-defined
since a left-multiplication by an element of  $K(\IR)$  only affects the 
lower triangular part of the decomposition, and it is injective since 
$P_{3}(\IC)\backslash G_3(\IC)  \cap M_3(\IR)$ consists only of the identity.
In particular, choosing $g=\Omega^{-1} e_Z \Omega$ 
where $e_Z$ is the coset representative in \eqref{Zef}, we
can express $(\mX,\mZ)$ as a function of the coordinates 
on the base $U,\tau^i,\bar\tau^i,\zeta^\Lambda,\tzeta_\Lambda,\sigma$ and the
complex coordinates $X_{IJ}$ on the fiber (the resulting expressions
turn out to be very cumbersome and are best omitted here). Note that $(\mX,\mZ)$
is independent of $\bar X_{IJ}$, as the two $\bar X$-dependent 
factors in \eqref{hatef} only affect the lower triangular part. Thus,
the Harish-Chandra embedding provides a holomorphic 
parametrization of the"twistor lines", i.e. the fibers of the projection $Z\to \cM_3$. This map
was also referred to as "the twistor map" in \cite{Neitzke:2007ke}. 

Conversely, an element of  $(\mX,\mZ) \in P_{3}(\IC)\backslash G_3(\IC)$ may be mapped 
into an element of $G_{3}(\IR)$
\begin{equation}
\label{eZtop}
e_{Z}=
\begin{pmatrix}\CI_4&0&0\\\bar{\mZ}&\CI_{n_v+2}&0\\
\bar{\mXt}&\bar{\mZ}^T  &\CI_4
\end{pmatrix}
\cdot
\begin{pmatrix}A^{-1/2}&0&0\\0&B^{-1/2}&0\\0&0&(A^T)^{1/2}
\end{pmatrix}
\cdot
\begin{pmatrix}\CI_4&\mZ&
\mXt\\0&\CI_{n_v+2}&\mZ^T\\0&0&\CI_4
\end{pmatrix}
\end{equation}
where $A$ and $B$ are $4\times 4$ and $(n_v+2)^2$ matrices
afforded by the decomposition
\be
\label{satso82}
\begin{pmatrix}\CI_4&\mZ&
\mXt \\0&\CI_{n_v+2}&\mZ^T\\0&0&\CI_4
\end{pmatrix}
\cdot
\begin{pmatrix}\CI_4&0&0\\\bar{\mZ}&\CI_{n_v+2}&0\\
\bar{\mXt} \bar{\mZ}&\bar{\mZ}^T  &\CI_4
\end{pmatrix}
=
\begin{pmatrix}*&&\\ *&*&\\ *&*&*\end{pmatrix} \cdot
\begin{pmatrix}A&0&0\\0&B&0\\0&0&A^{-T}
\end{pmatrix}\cdot
\begin{pmatrix}*&*&*\\ &*&*\\ & &*\end{pmatrix}\ .
\ee
The Iwasawa decomposition of $e_Z$ then allows to express the coordinates
$U,\tau^i,\bar\tau^i,\zeta^\Lambda,\tzeta_\Lambda,$ $\sigma,X,\bar X$ on $\cM_3 \times F$
in terms of $(\mX,\mZ)$ and their complex conjugates. This reciprocal map was termed
"covariant c-map", or superconformal quotient, in \cite{Neitzke:2007ke}. Again, this
map is in principle computable, but the resulting expressions are too cumbersome
to be of any practical use.

While only the real group $G_3(\IR)$ acts on the base $\cM_3$, the action on the
twistor space $Z$ can be extended to the complexified group  $G_3(\IC)$: it acts
by right-multiplication on the coset representative \eqref{truehc}, followed by 
a left-multiplication by an appropriate lower triangular matrix so as to return to
the strictly upper triangular gauge.  The complex coordinates $(\mX,\mZ)$  are
adapted to the holomorphic action of the nilpotent group of strictly
upper-block diagonal matrices, in the sense that no compensating
left-action is needed. This action is generated by the vector fields
\be 
E^{a}_I = \pa_{\mZ^{I a}} + \eps_{IJKL} \mZ^{J a} \pa_{\mX_{KL}}\ ,\quad 
E^{IJ} =    \pa_{\mX_{IJ}}
\ee
which satisfy the Heisenberg-type commutation relations
\be
\left[ E^{a}_I, E^{b}_J \right] =  \eps_{IJKL} \, \delta_{ab}\, E^{KL} \ .
\ee
For applications to black hole physics, it would be desirable to 
have complex coordinates adapted to the Heisenberg algebra \eqref{heisr}, 
which corresponds to the electric, magnetic and NUT charges. As for 
the  $SU(2,1)$ case studied in \cite{Gunaydin:2007qq}, it should be possible to 
obtain this change of variable by taking the limit $U\to -\infty, \tau_2\to 0$
in the twistor map. We note however that for $n_v=0$, there is an obvious holomorphic action
of $G_3(\IR)$ on $14$ complex variables, adapted to Heisenberg algebra \eqref{heisr},
corresponding to the "fake" Harish-Chandra decomposition in the original
basis \eqref{eta262}, 
\be
\begin{pmatrix} 
* &&&&\\
*&* &&&\\
*&*&*&&\\
*&*&*&*&*\\
*&*&*& &*
\end{pmatrix}
\cdot
\begin{pmatrix} 
1&\beta & \xi &   -\frac12 \xi \xi^t &\alpha-\frac12 \xi \txi^t  \\
0& 1&\txi & -\alpha-\frac12 \xi \txi^t &  -\frac12 \txi \txi^t\\
&&\CI_{6} & -\xi^t & -\txi^t \\
&&&1& 0 \\
&&&-\beta&1
\end{pmatrix}
\label{fakehc}
\ee
where $\alpha$ and $\beta$ are two complex variables, and 
$\xi^\Lambda$, $\txi_\Lambda$ are two complex vectors in $\IC^{6}$. 
It would be interesting to find the change of variable from the complex
coordinates $(\mX,\mZ)$ to $(\xi^\Lambda, \txi_\Lambda, \alpha,\beta)$.

An Hermitian metric on $Z$ can be obtained by computing the
right-invariant form, projecting out the $M_3(\IC)$ part, and 
taking $M_3$-invariant quadratic combinations as in \eqref{dsnu}. 
The strictly upper-triangular components of $dg\cdot g^{-1}$
provide right-invariant (1,0) forms
\be
\label{th10m}
\theta_{IJ}^{Z} = \tilde V ( d\mX + \frac12 \mZ d\mZ^T - \frac12 d\mZ \mZ^T ) \tilde V^T\ ,
\quad \theta_{I a}^Z= \tilde Vd\mZ \ .
\ee
where
\be
\tilde V = A^{-1/2}
\ee
Setting $\tilde V=1$  in \eqref{th10m}, we obtain a basis of holomorphic (i.e. $\bar\pa$-closed)
(1,0) forms,
\be
\label{DXDZ}
D\mX \equiv d\mX + \frac12 \left( \mZ d\mZ^T - d\mZ \mZ^T \right)\ ,\quad DZ= d\mZ\ .
\ee
The antisymmetric matrix of holomorphic 1-forms $D\mX$ plays the r\^ole
of the holomorphic contact distribution in the \qk case. Note that under a 
right-action of $G(\IC)$, $D\mX_{IJ}$ transforms by an element of $GL(4,\IC)$,
corresponding to the block diagonal component of the compensating 
lower triangular matrix required to restore the upper triangular gauge. 
In the \qk case, this issue can be circumvented by introducing a new
$\IC^\times$ valued variable $t$, and considering the one-form
$t D\mX$: the rescaling of $DX$ can be reabsorbed by a rescaling
of $t$, leading to a globally defined holomorphic one-form
on $\IC^\times \times Z$, whose exterior derivative is the holomorphic two-form $\Omega$
on the hyperk\"ahler cone of $\cM_3$ \cite{de Wit:2001dj}. In the present
case, one may similarly introduce 6 new variables $t^{IJ}$ and consider
the globally defined holomorphic two-form $\Omega=d( t^{IJ} X_{IJ})$.
We shall return to this possibility momentarily.

In the above construction of the metric \eqref{dsnu},  it is difficult to fix
the coefficient $\nu$ such that the metric is K\" ahler. However, 
according to the general prescription of \cite{Satake}, we know that
 a K\"ahler potential for an invariant metric on $Z$ is given by the logarithm 
of a character of $M_3(\IC)$ evaluated on the block diagonal part in the
decomposition \eqref{satso82}\footnote{Note that $K_Z$ reduces to the K\"ahler 
potential \eqref{KX} on the fiber $F$ at $\mZ=0$. }:
\be
\label{KZ}
K_Z(\mX,\mZ,\bar{\mX},\bar{\mZ}) = \log \det \left[ \CI_4 + \mZ\bar \mZ+ \mXt \bar{\mXt}^T \right]
\ee
Comparison to the metric \eqref{dsnu} fixes $\nu=-1$. It would be interesting to check
whether  the metric is K\"ahler-Einstein, as in the case of twistor spaces of \qk spaces.

Given the transformation properties of the
kernel matrix $A(\mX,\mZ,\bar{\mX},\bar{\mZ})$, it is also natural to consider 
higher dimensional spaces with K\"ahler potential
\be
K(t,\mX,\mZ,\bar t,\bar{\mX},\bar{\mZ}) = \bar t \cdot R\left[  \CI_4 + \mZ\bar \mZ+ \mXt \bar{\mXt}^T
 \right] \cdot t
\ee
where $t$ transforms in some finite dimensional representation $R$ of $GL(4,\IC)$. 
For $t=t^{IJ}$ in the 6-dimensional antisymmetric representation, combining
this result with the construction in the paragraph below \eqref{DXDZ},  we obtain a 
$8(n_v+5)$ real-dimensional K\"ahler space with a (2,0)  holomorphic form and
a homothetic Killing vector. It is natural to conjecture that this provides 
the natural hyperk\"ahler metric \cite{kronheimer} on a 
complex nilpotent co-adjoint orbit of $SO(n_v+10,\IC)$ 
associated to the partition $(2^4,1^{n_v+2})$, of
complex dimension $4(n_v+5)$.

\subsection{Supersymmetry and holomorphy}
We now return to the physical motivation for this geometric construction,
the supersymmetry conditions \eqref{susycond}. As we discussed below
\eqref{ga}, there is an equivalence
\be
 \vareps_\mu  \,\Gamma^\mu_{A' A} \, p^{Aa}  = 0 \quad \Leftrightarrow \quad 
 ( \CI_4 | X )_{IA} \,p^{Aa} = 0\ ,
\ee
provided the null vector $\vareps_\mu$ is related to $X_{IJ}$ via \eqref{epsy},
\eqref{XtoY}. Moreover, in \eqref{th10s}, we have established that the one-forms
$DZ_{I a}=( \CI_4 | X )_{IA} p^{Aa}$ are (1,0) forms with respect to the complex
structure on $Z$. Therefore, if we lift the geodesic motion on $\cM_3$ to the twistor 
space $Z$ by requiring that at every point, $DX_{IJ}=0$, we conclude that 
supersymmetric geodesics on $\cM_3$ have a tangent vector of type (0,1)
at every point, and therefore correspond to an anti-holomorphic curves $\rho: \IC \to Z$.
In practice, this means that the holomorphic coordinates $z^i$ 
are constant along
the flow, while the anti-holomorphic coordinates   $\bar z^{\bar i}$  
evolve\footnote{Due to the analytic continuation
from $\cM_3$ to $\cM_3^*$, the complex coordinates $z^i$
should be treated as independent variables.} in such a way that the gradient
of the K\"ahler potential grows linearly with the affine 
parameter~\cite{Neitzke:2007ke}\footnote{This follows directly from the geodesic equation
$\ddot z^i + \Gamma^i_{jk} \dot z^j \dot z^k=0$, given that the Christoffel symbol $\Gamma^i_{jk}$
has no mixed holomorphic/anti-holomorphic components.}
\be 
\pa_{z^i} K = c_i \tau + d_i\ .
\ee
Moreover, the BPS constraints \eqref{p2}, re-expressed as $DZ_{Ia}=DX_{IJ}=0$, now manifestly 
form a system of first class constraints, as the Lie bracket of two (anti)holomorphic vectors
is necessarily (anti)holomorphic. As in the $\cN=2$ case \cite{Gunaydin:2007bg}, we can
therefore identify the 1/4-BPS phase space as the twistor space  $Z$, equipped
with its K\" ahler form. 

In order to make the best use of this geometric statement, it would be desirable
to construct a coordinate system on $Z$ adapted to the Heisenberg 
symmetries \eqref{heisr}. This would enable us to determine the most general 
1/4-BPS spherically symmetric solutions in $\cN=4$ supergravity, and also
to compute the exact BPS black hole wave function as a Penrose transform of a 
holomorphic wave-function on $Z$,  along the lines
 of \cite{Neitzke:2007ke}. While we have been unable to carry out this computation,
 in the next subsection we construct some holomorphic functions on $Z$ which 
 provide BPS wave functions for solutions with certain charges.

\subsection{Some holomorphic functions on $Z$}

In this section, we construct some holomorphic functions on $Z$ in the
coordinate system $U,\tau^i,\bar\tau^i,\zeta^\Lambda,\tzeta_\Lambda,\sigma,X,\bar X$ 
adapted to the fibration $F \to Z \to \cM_3$. For ease of notation, we denote the entries 
in $X_{IJ}$ as
\be
X = \left(
\begin{array}{llll}
 0 & y_1 & y_2 & y_3 \\
 -y_1 & 0 & x_3 & -x_2 \\
 -y_2 & -x_3 & 0 & x_1 \\
 -y_3 & x_2 & -x_1 & 0
\end{array}
\right)
\ee
and similarly for $\bar X$. 

Our first observation is that $x_1,x_2,x_3$ and $y_2/y_1$, $y_3/y_1$ are holomorphic
functions on $Z$. This follows from the fact that their differentials are of (1,0) type,
\bea
dx_1 &=& Dx_1+\frac{1}{\sqrt{2}}\left(-y_3 DZ_{31}+i y_3 DZ_{32}+y_2
   DZ_{41}-i y_2 DZ_{42}\right)\\
dx_2 &=&   Dx_2+\frac{1}{\sqrt{2}}\left(y_3
   DZ_{21}-i y_3 DZ_{22}-y_1 DZ_{41}+i y_1
   DZ_{4,2}\right)\nn\\
dx_3 &=& Dx_3+\frac{1}{\sqrt{2}} \left(-y_2 DZ_{21}+i y_2
   DZ_{22}+y_1 DZ_{31}-i y_1 DZ_{32}\right)\nn
   \eea
   \bea
d\left(\frac{y_2}{y_1}\right) &=&    
   \frac{2 y_1 Dy_2 -2  y_2 Dy_1+\sqrt{2} (y_2 (DZ_{21}+i
   DZ_{22})-y_1 (DZ_{31}+i DZ_{32}))}{2 y_1^2}\nn\\
   d\left(\frac{y_3}{y_1}\right)  &=&    
   \frac{2  y_1 Dy_3 -2  y_3 Dy_1+\sqrt{2} (y_3 (DZ_{21}+i
   DZ_{22})-y_1 (DZ_{41}+i DZ_{42}))}{2 y_1^2}\nn
\eea
Secondly, we note that the contraction of any Killing vector $\kappa^m \pa_{\phi^m}$ 
with the holomorphic 
contact distribution $D\mX_{IJ}$ yields a $4\times 4$ antisymmetric matrix of holomorphic
functions, since $G_3(\IR)$ acts holomorphically on $Z$. Moreover the one forms,  
$D\mX_{IJ}$ and $DX_{IJ}=DX_{IJ,m}  d\phi^m$ 
are related to each other by a $GL(4,\IC)$ transformation,
\be
DX = V^{-1}  \tilde V \cdot D\mX\cdot \tilde V^T V^{-T}\ .
\ee 
Thus, for two Killing vectors $\kappa^m$ and $\kappa'^m$, the combination
\be
\langle \kappa, \kappa' \rangle \equiv \,\epsilon^{IJKL} \kappa^m DX_{IJ,m} \,\kappa'^n DX_{KL,n}
\ee
 is holomorphic, up to an overall factor independent of $\kappa$ and $\kappa'$. It may be checked
 explicitly that the product
 \be
 y_1^{-1} e^{-2U}  \, \langle \kappa, \kappa' \rangle
 \ee
 is holomorphic for one choice  of $\kappa$ and $\kappa'$, and therefore for any pair of Killing
vectors. Different pairs $(\kappa,\kappa')$ may not necessary give independent holomorphic
functions however: for $n_v=0$, an explicit computation shows that a linear basis of holomorphic
functions obtained in this way, using the Killing vectors $P^\Lambda, Q_\Lambda, K,
H, Y_0, Y_\pm$ introduced in Section \ref{1d}, may be chosen as 
\be
\label{basehol}
\langle P^\Lambda, H \rangle\ ,\quad \langle Q_\Lambda, H \rangle,\quad
\langle P^\Lambda, Q_\Sigma\rangle = - \langle Q_\Lambda, P^\Sigma\rangle \ ,\quad
\langle H, H \rangle \ .
\ee
The remaining non-vanishing inner products can be expressed in terms of this basis as
\be
\langle P^\Lambda, Y_0 \rangle = \langle P^\Lambda, H \rangle\ , \quad
\langle Q_\Lambda, Y_0 \rangle = -\langle Q_\Lambda, H \rangle\ ,\quad
\langle Y_0, Y_0 \rangle = \langle Y_+, Y_- \rangle =   -\langle H, H \rangle 
\ee
This provides non-trivial examples of holomorphic functions on $Z$. Unfortunately,
we have not managed to find eigenfunctions of the charge generators $P^\Lambda,
Q_\Lambda, K$. Instead, one may check that the action of the Killing vectors on 
the holomorphic functions \eqref{basehol} is given by
\bea
P^\Lambda &\cdot& \langle P^\Sigma, H \rangle = 0\ ,\quad
Q_\Lambda \cdot \langle Q_\Sigma, H \rangle = 0 \ ,\quad
K \cdot \langle P^\Lambda, H \rangle = 0\ ,\\
K &\cdot& \langle Q_\Sigma, H \rangle = 0 \ ,\quad
Y_+ \cdot \langle P^\Lambda, H \rangle =  0 \ ,\quad
Y_+ \cdot \langle Q_\Sigma, H \rangle = 0\ , \\
P^\Lambda &\cdot& \langle Q_\Sigma, H \rangle =    \langle P^\Lambda, Q_\Sigma  \rangle \ ,\qquad
Q_\Lambda \cdot \langle P^\Sigma, H \rangle = -\langle P^\Lambda, Q_\Sigma  \rangle \ .
\eea

\section{Discussion}

In this work, we have analyzed $1/4$-BPS spherically symmetric, stationary configurations in $D=4$, 
$\cN=4$ supergravity, by dimensional reduction to one (radial) dimension. In parallel with the
treatment of BPS black holes in $D=4, \cN=2$ supergravity \cite{Neitzke:2007ke,Gunaydin:2007bg},
we have shown that such configurations correspond to supersymmetric geodesics on 
the three-dimensional symmetric moduli space $\cM_3$. This provides a powerful technique for obtaining new black hole solutions in 4 dimensions. Indeed, we have found that the phase
space of BPS solutions is given by a degree 3 nilpotent orbit in $SO(8,2+n_v)$, whose real 
dimension $8n_v+28$ is  twice as large as expected by extrapolating the results for $\cN=2$  
black holes. We have also found indications that the phase space of non-BPS extremal black holes 
is given by a nilpotent orbit with the  same complexification as in the BPS case, but related by 
an outer automorphism of the real group $SO(8,2+n_v)$. It would be interesting to study this further.

In addition, we have shown that supersymmetric geodesics on $\cM_3$
can be lifted to holomorphic curves on a
homogeneous complex space $Z$, the twistor space \eqref{defZ}. In contrast to the 
$\cN=2$ case, the fiber does not parametrize the sphere
of complex structures $S^6$, but rather the space $SO(8)/U(4)$ of isotropic 4-planes in $\IC^8$.
Moreover, $Z$ does not carry a holomorphic contact form, but rather a $4\times 4$ 
antisymmetric matrix of holomorphic contact forms. This complication has so far prevented
us from constructing complex coordinates adapted to the Heisenberg symmetries of the 
problem, which were instrumental in \cite{Neitzke:2007ke,Gunaydin:2007bg} for obtaining
the BPS radial wave function for a black hole with fixed electric and magnetic charges.
Nevertheless, there is no doubt that such a system can be constructed, and that a Penrose-type correspondence can be set up between holomorphic functions on $Z$ and solutions of the 
second order partial differential equation 
$(C_{AB} \nabla^{Aa} \nabla^{Bb} + \lambda \delta^{ab})\Psi=0$, which 
follows by quantizing \eqref{p2}.  Irrespective of applications to black hole physics, this 
correspondence may be used to compute instanton corrections in 3 dimensions,
provided one can identify a coupling in the low energy effective action governed by 
the same partial differential equation.

\acknowledgments
 B. P.  is grateful to A. Neitzke, S. Vandoren and A. Waldron for discussions on 
 related twistor constructions for \qk symmetric spaces.
The research of B.P. is supported in part by ANR(CNRS-USAR)
contract no.05-BLAN-0079-01. 

\appendix

\section{$SO(8)$ Gamma Matrices  \label{so8gamma}}
In this section, we describe our conventions for the $SO(6)$ and
$SO(8)$ Dirac matrices used in the text. We start with 
the $4\times 4$ $SO(6)$ Sigma matrices $\Sigma_i$ ($i=1\dots 6$)
\be
\Sigma_1 = 
\left(
\begin{array}{llll}
 0 & 1 & 0 & 0 \\
 -1 & 0 & 0 & 0 \\
 0 & 0 & 0 & 1 \\
 0 & 0 & -1 & 0
\end{array}
\right)\ ,\quad
\Sigma_2 = 
\left(
\begin{array}{llll}
 0 & i & 0 & 0 \\
 -i & 0 & 0 & 0 \\
 0 & 0 & 0 & -i \\
 0 & 0 & i & 0
\end{array}
\right)
\ ,\quad
\Sigma_3 = 
\left(
\begin{array}{llll}
 0 & 0 & 1 & 0 \\
 0 & 0 & 0 & -1 \\
 -1 & 0 & 0 & 0 \\
 0 & 1 & 0 & 0
\end{array}
\right)
\ ,\quad
\ee
\be
\Sigma_4 = 
\left(
\begin{array}{llll}
 0 & 0 & i & 0 \\
 0 & 0 & 0 & i \\
 -i & 0 & 0 & 0 \\
 0 & -i & 0 & 0
\end{array}
\right)
\ ,\quad
\Sigma_5 = 
\left(
\begin{array}{llll}
 0 & 0 & 0 & 1 \\
 0 & 0 & 1 & 0 \\
 0 & -1 & 0 & 0 \\
 -1 & 0 & 0 & 0
\end{array}
\right)
\ ,\quad
\Sigma_6 = 
\left(
\begin{array}{llll}
 0 & 0 & 0 & i \\
 0 & 0 & -i & 0 \\
 0 & i & 0 & 0 \\
 -i & 0 & 0 & 0
\end{array}
\right)
\ ,\quad
\ee
corresponding to the $SO(6)$ Dirac matrices in the Weyl representation,
\be
\Gamma_i = \begin{pmatrix}0 & \tilde\Sigma_i \\ \Sigma_i & 0
\end{pmatrix}\ ,\quad
\{\Gamma_i,\Gamma_j\}=2\delta_{ij} \CI_8
\ee
where $\tilde\Sigma_i = (-1)^{i} \Sigma_i= \Sigma_i^\dagger$. This is extended to a
representation of the Clifford algebra of $SO(7)$ by adding 
$\Gamma_7 = i\Gamma_1\Gamma_2\Gamma_3\Gamma_4\Gamma_5\Gamma_6$.
The charge conjugation matrix is given by
\begin{equation}
C = -i\Gamma_2\Gamma_4\Gamma_6 = \begin{pmatrix} 0 & \CI_4 \\
\CI_4 & 0 \end{pmatrix}\ ,\quad C \Gamma_i^T C^{-1} = -\Gamma_i \ .
\end{equation}
The matrices $\Gamma_i, \Gamma_7$ supplemented with $\Gamma_0=i\CI_8$,
can then serve as Sigma matrices\footnote{We keep the same symbol $\Gamma$
to avoid unnecessary extra notation.} $\Gamma^{\mu}_{A'A}$ ($\mu=0\dots 7$) 
for a chiral representation of the
Clifford algebra of $SO(8)$. In particular, the Lorentz generators $\Gamma^{\mu\nu}_{AB}$ in the 
spin $8_S$ representation of $SO(8)$ are given by 
\begin{equation}
\Gamma^{\mu\nu} = [\Gamma^{\mu}, \Gamma^{\nu}]\ ,\quad
\Gamma^{\mu 0} = 2i\Gamma^{\mu}, \quad
\Gamma^{0 \mu} = -2i\Gamma^{\mu} \quad (\mu,\nu \neq 0)
\end{equation}
satisfying the $SO(8)$ algebra,
\begin{equation}
\left[\Gamma^{\mu\nu},\Gamma^{\rho\sigma}\right]=-4\left(
\delta^{\mu\rho} \Gamma^{\nu\sigma} + \delta^{\nu\sigma} \Gamma^{\mu\rho}
-\delta^{\nu\rho} \Gamma^{\mu\sigma} -\delta^{\mu\sigma} \Gamma^{\nu\rho} \right)
\end{equation}
Similarly, the Lorentz generators $\tilde\Gamma^{\mu\nu}_{AB}$ in the 
spin $8_C$ representation of $SO(8)$ can be constructed as 
\be
\tilde\Gamma_i = \begin{pmatrix}0 & \Sigma_i \\ \tilde\Sigma_i & 0
\end{pmatrix}\ ,\quad 
\tilde\Gamma_0 = i \CI_8\ ,\quad
\tilde\Gamma_7 = i \tilde\Gamma_1\tilde\Gamma_2\tilde\Gamma_3
\tilde\Gamma_4\tilde\Gamma_5\tilde\Gamma_6
\ee
\be
\tilde C = -i\tilde\Gamma_2\tilde\Gamma_4\tilde\Gamma_6 = \begin{pmatrix} 0 & \CI_4 \\
\CI_4 & 0 \end{pmatrix}
\ee
\begin{equation}
\tilde\Gamma^{\mu\nu} = [\tilde\Gamma^{\mu}, \tilde\Gamma^{\nu}]\ ,\quad
\tilde\Gamma^{\mu 0} = 2i\tilde\Gamma^{\mu}, \quad
\tilde\Gamma^{0 \mu} = -2i\tilde\Gamma^{\mu} \quad (\mu,\nu \neq 0)
\end{equation}

We note that the triality automorphism is implemented by taking an
antisymmetric matrix $\Omega_{IJ}^V$ in the vector representation to
matrices $\Omega_{AB}^S$ or  $\Omega_{A'B'}^C$ in the spinor representation via
\begin{equation}
\Omega_{\mu\nu}^V \, \Gamma_{AB}^{\mu\nu} = \Omega_{AB}^S \ ,\quad
\Omega_{\mu\nu}^V \, \tilde\Gamma_{A'B'}^{\mu\nu} = \Omega_{A'B'}^C \ ,
\label{trialitymap}
\end{equation}

\section{Nilpotent orbits in orthogonal groups}

In this appendix, we briefly review some general facts about nilpotent co-adjoint orbits,
before restricting to orthogonal groups. The proofs of all these results 
can be found in~\cite{Collingwood}.

Complex nilpotent orbits in $G$ are classified by conjugacy classes of 
homomorphisms $\mathfrak{su}(2) \to \mathfrak{g}$, i.e. 
triplets $e,f,h$ of elements in the Lie algebra $\mathfrak{g}$ of $G$
satisfying the $SU(2)$ algebra, $[e,f]=h, [h,e]=2e, [h,f]=-2f$. Under the adjoint action of 
this $SU(2)$, $\mathfrak{g}$ decomposes into a sum of finite-dimensional 
representations.  $\mathfrak{g}$ 
may be further decomposed as a sum of eigenspaces of the Cartan generator 
$h$,  $\mathfrak{g}=\sum_{i=-i_0\dots i_0} \mathfrak{g}_i$.
The complex nilpotent orbit is isomorphic to $P\backslash G$,
where $P$ is obtained by exponentiating $\mathfrak{p}=\sum_{i\leq 0} \mathfrak{g}_i$.
The real dimension of the nilpotent orbit is given by $\dim \mathfrak{g} 
- \dim \mathfrak{g}_0 - \dim \mathfrak{g}_1= 2 (\dim \mathfrak{g} - \dim \mathfrak{p})$.  
The set of all nilpotent orbits admits a partial ordering,  the closure ordering,
whereby $e<e'$ if $e'$ lies in the closure of the nilpotent orbit through $e'$.
All nilpotent orbits of a given group $G$ can be displayed in a Hasse-type diagram, 
with vertically increasing dimensions and links corresponding to the closure ordering. 

For $G=GL(N,\IC)$, complex nilpotent orbits
are in one-to-one correspondence with partitions of $N$,
i.e. Young tableaux with $N$ boxes. The partition corresponds to the 
Jordan normal form of the nilpotent element $e$, or to the dimensions of
the representations appearing in the decomposition of $\mathfrak(g)$
under $SU(2)$. For $G=SO(N,\IC)$, 
complex nilpotent orbits are in one-to-one correspondence  with Young tableaux with $N$ boxes 
such that lines of even length always occur in pairs.  When $N$ is even, "very even"
partitions, corresponding to configurations with only rows of even length, are an 
exception to this rule, as they label two distinct orbits.
For $G=GL(N,\IC)$ or $G=SO(N,\IC)$, the closure ordering $e\leq e'$ holds
whenever for all $p=1\dots N$, the number of boxes in the first $p$ columns of the Young
Tableau associated to $e$ is less than the number of boxes in the first $p$ columns of the Young
Tableau associated to $e'$ (see \cite{Bachas:2000dx} for a physical realization 
of this ordering). 

In Table \ref{taborb}, we list the complex nilpotent orbits of 
$G=SO(n_v+10,\IC)$ whose dimension scales as $k n_v$ with $k\leq 8$ when $n_v\to\infty$,
\footnote{It is easy to see that the numbers of nilpotent orbits whose 
dimension scales as $k n_v$ is given by 
the coefficient of $q^k$ in the Taylor expansion of $1/\prod_{n=0}^{\infty}(1-q^{4n+2})
(1-q^{2n+2})$ around $q=0$.}; their closure relations are displayed in  the 
Hasse diagram in Figure \eqref{figorb}.
The table reveals two complex nilpotent orbits whose
real dimension equals the real dimension $8n_v+28$ of the twistor space $Z$.
The nilpotent orbit  $(5,2^4,1^{n_v-3})$  corresponds to a weight decomposition 
ranging from $i=-6$ to $i=6$ and bears no relation with $Z$. In contrast, the 
nilpotent orbit $(3^4,1^{n_v-2})$ gives rise to the same 5-grading as in  \eqref{comp5g},
\be
\label{comp5g2}
\mathfrak{g}=
6\vert_{-4} \oplus 4(n_v+2) \vert_{-2}\oplus \frac12(n_v^2+3n_v+34)\vert_{0} \oplus 4(n_v+2)\vert_{2}  \oplus 6\vert_{4}
\ee
and so is identical to the twistor 
space $Z$ \eqref{defZ}.  It may be worthwhile noting that the orbit $(2^4,1^{n_v+2})$ 
yields the same grading, but with half the charge; as a result its dimension is 
smaller by $4(n_v+2)$.
On the other hand, the orbit $(3^2,1^{n_v+4})$, of real dimension $4n_v+26$, gives the same
5-grading as \eqref{real5g},
\be
\mathfrak{g}=1\vert_{-4} + (2n_v+12)\vert_{-2} + \frac12(n_v^2+11n_v+38)\vert_{0} 
+ (2n_v+12)\vert_{2} \oplus 1\vert_{4}
\ee
which is adapted to the complex structure on the twistor space of the 
\qk manifold $SO(4,n_v+6)/SO(4)SO(n_v+6)$. Again, the orbit  $(2^2,1^{n_v+6})$ 
gives the same grading but with half the charge.
Finally, the orbit $(3,1^{n_v+7})$ of real dimension $2(n_v+8)$ gives a three-grading
\be
\mathfrak{g}=(n_v+8)\vert_{-2}\oplus \frac12(n_v^2+15 n_v+58)\vert_{0}  \oplus  (n_v+8)\vert_{2}
\ee
adapted to the complex structure on $SO(2,n_v+8)/SO(2)\times SO(n_v+8)$.

\EPSFIGURE[h]{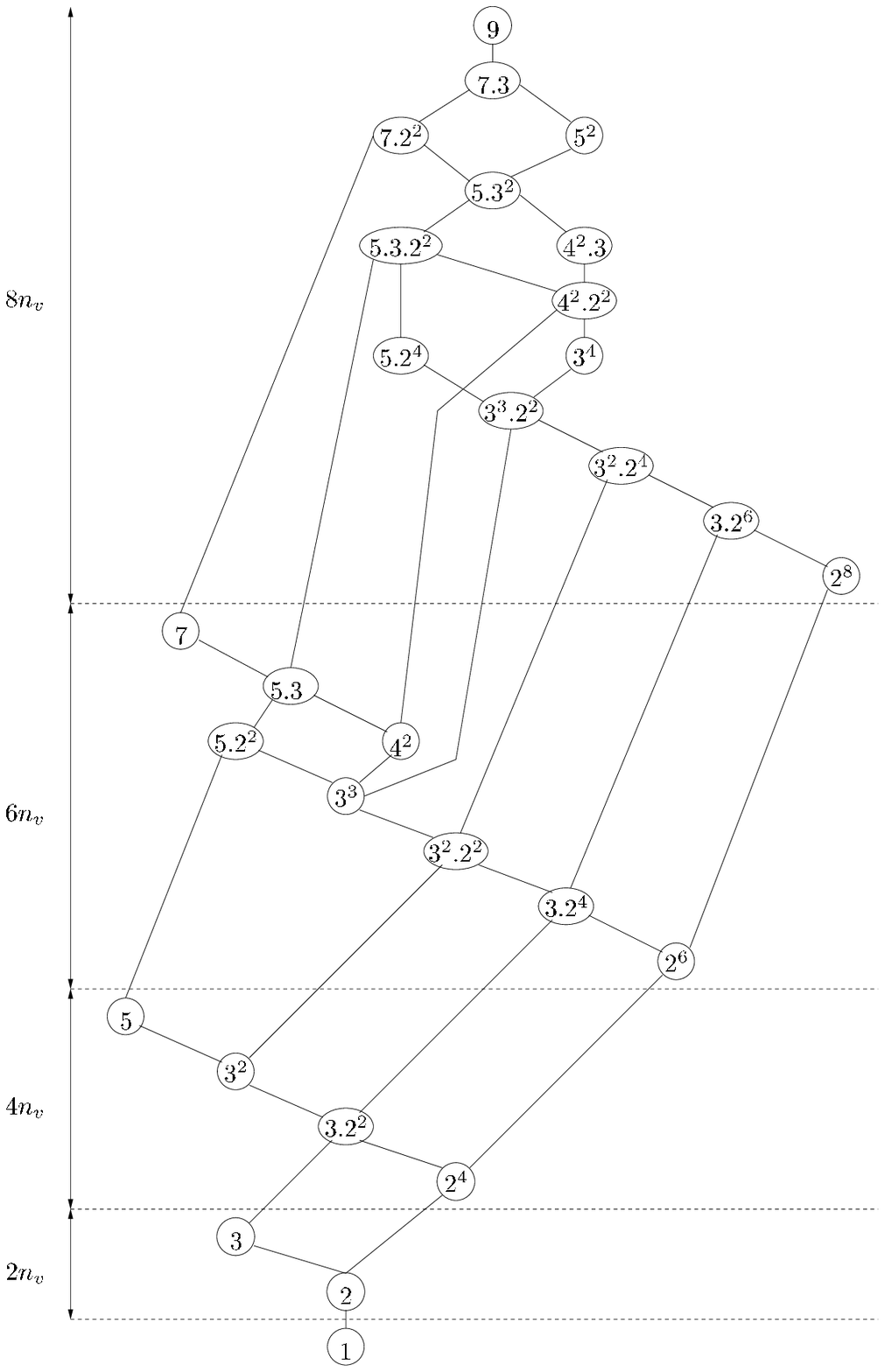, height=18cm}{Hasse diagram for the nilpotent orbits in Table 1.\label{figorb}}

\begin{table}
$$
\begin{array}{|@{\hspace*{5mm}}c@{\hspace*{5mm}}|@{\hspace*{5mm}}c@{\hspace*{5mm}}|}
\begin{array}{lcl}
  (2^2, 1^{n_v+6}) &:& 2 n_v+14\\ 
  (3,1^{n_v+7})&:& 2 n_v+16\\ 
  \\
  (2^4,1^{n_v+2})&:& 4 n_v+20\\
  (3,2^2,1^{n_v+3})&:& 4 n_v+24\\
  (3^2,1^{n_v+4})&:& 4 n_v+26\\
  (5,1^{n_v+5})&:& 4 n_v+28\\ 
  \\
  (2^6,1^{n_v-2})&:& 6n_v+18\\
  (3,2^4,1^{n_v-1})&:& 6 n_v+24\\
  (3^2,2^2,1^{n_v})&:& 6 n_v+28\\
  (3^3,1^{n_v+1})&:& 6 n_v+30\\
  (4^2,1^{n_v+2})&:& 6 n_v+32\\
  (5,2^2,1^{n_v+1})&:& 6 n_v+32\\
  (5,3,1^{n_v+2})&:& 6 n_v+34\\
  (7,1^{n_v+3})&:& 6 n_v+36\\
\end{array}
&
\begin{array}{lcl}
  (2^8,1^{n_v-6})&:& 8n_v+8\\
  (3,2^6,1^{n_v-5})&:& 8n_v+16\\
  (3^2,2^4,1^{n_v-4})&:& 8 n_v+22\\
  (3^3,2^2,1^{n_v-3})&:& 8 n_v+26\\
  (3^4,1^{n_v-2})&:& 8 n_v+28\\
  (5,2^4,1^{n_v-3})&:& 8n_v+28\\
  (4^2,2^2,1^{n_v-2})&:& 8n_v+30\\
  (4^2,3,1^{n_v-1})&:& 8n_v+32\\
  (5,3,2^2,1^{n_v-2})&:& 8n_v+32\\
  (5,3^2,1^{n_v-1})&:& 8n_v+34\\
  (7,2^2,1^{n_v-1})&:& 8n_v+36\\
  (5^2,1^{n_v})&:& 8n_v+36\\
  (7,3,1^{n_v})&:& 8 n_v+38\\
  (9,1^{n_v+1})&:& 8n_v+40\\
\end{array}
\end{array}
$$
\caption{Complex nilpotent orbits in $SO(n_v+10,\IC)$ with 
dimension $\mathcal{O}(k n_v)$ with $k\leq 8$. \label{taborb}}
\end{table}

We now turn to the classification of real nilpotent orbits, which is rather more subtle. 
For real orthogonal groups $SO(p,q,\IR)$,  nilpotent orbits
are classified by Young tableaux with $N=p+q$ boxes as above, with
additional assignments of a sign $\pm$ to each  box such that signs alternate along 
lines, rows of even length start with $+$, and the total number of (plus,minus) signs
is $(p,q)$.  A given signed Young tableau may corresponds to 4 different 
orbits when all rows have even length, 2 different orbits when all rows with
odd length have an even number of $+$, 2 different orbits when all rows with
odd length have an even number of $-$, and a unique orbit in other cases \cite{Collingwood}.
For example, $SO(8,2,\IR)$ admits 7 non-zero nilpotent real orbits, corresponding to the partitions
\be
(2^2,1^6)\ ,\quad (3,1^7)_{I,II,III}\ ,\quad (3^2,1^4)\ ,\quad (5,1^5)_{I,II}\ . 
\ee
of dimension 14, 16, 26 and 28 and nilpotency degree 2,3,3,5, respectively. 
In particular, there are 4 inequivalent nilpotent orbits of degree 3, none of
whose dimension agrees with the dimension $Z$ (the orbits $(5,1^5)_{I,II}$
do happen to have dimension 28, but are related to a 13-th grading, 
as indicated  above). This is an artifact of this low-rank case, since
the nilpotent orbit $(3^4,1^{n_v-2})$ does appear in the list of 
real nilpotent orbits of $SO(8,2+n_v,\IR)$ for $n_v\geq 2$. Choosing
the sign configuration $[(+-+)^4,(-)^{n_v-2}]$, all rows have odd length
and carry an odd number of minuses, so this configuration appears
in two varieties, related by an outer automorphism of $SO(8,n_v+2)$.

\end{document}